\documentclass{jpp}
\usepackage{graphicx}
\usepackage{multirow}
\usepackage[utf8]{inputenc}
\usepackage[T1]{fontenc}
\usepackage{amsmath}
\usepackage{xcolor}

\usepackage{verbatim}
\usepackage{physics}
\usepackage{ulem}
\usepackage{hyperref}
\usepackage{float}
\usepackage{graphicx}

\newcommand{\eq}[1]{equation~(\ref{#1})}
\newcommand{\eqs}[2]{equations~(\ref{#1}--\ref{#2})} 

\shorttitle{tearing vs. Weibel}
\shortauthor{K. M. Schoeffler et al.}

\title{Are the tearing and the Weibel instabilities the same?}

\author{K. M. Schoeffler\aff{1}
  \corresp{\email{Kevin.Schoeffler@rub.de}},
  H. Aravindakshan\aff{1},
  F. Pucci\aff{2},
  M. E. Innocenti\aff{1}}

\affiliation{\aff{1}Institut f\"ur Theoretische Physik, Ruhr-Universit\"at Bochum, Bochum, Germany\\
\aff{2}Jet Propulsion Laboratory, Pasadena, CA 91109, USA}

\begin{document}

\maketitle

\begin{abstract}
The tearing instability, which extracts free energy from a magnetic field gradient, and the Weibel instability, which draws free energy from temperature anisotropies, appear at first glance to be fundamentally distinct.
However, the magnetic field gradient in the tearing instability requires a current, and the resulting plasma drift leads to an effective temperature anisotropy, so that the plasma is hotter along the direction of the current flow. This anisotropy can play an essential role in driving the instability. We demonstrate, using semi-implicit particle-in-cell simulations, a smooth transition between the tearing regime for thick current sheets, where the anisotropy has little influence, and a thin-current-sheet regime better described by Weibel physics. We find that for current sheet thicknesses less than the ion Larmor radius, where standard assumptions of the tearing instability break down, the growth rate of the mode commonly identified as the tearing instability agrees remarkably well with the growth rate predicted by ion-cyclotron theory, which is essentially the magnetized counterpart of the Weibel instability.
\end{abstract}

\section{Introduction}\label{sec:intro}

The tearing instability~\citep{Furth1963} is driven by free energy contained in     
opposite-directed magnetic fields that are separated by a thin current sheet (in comparison to the system size).
While this energy is eventually released in the non-linear process of magnetic reconnection, which converts the free energy into more bulk flows, plasma heating, and non-thermal high-energy particles, the tearing instability is associated with the initial linear stage of this process. A variant of this instability, known as the plasmoid instability, can also occur in the nonlinear stage of reconnection when it is fast compared to the reconnection time scales \citep{Loureiro2007,Bhattacharjee2009}. Although originally a fluid theory, kinetic models of the tearing instability were also developed \citep{Laval1966,Coppi1966,Zelenyi1979}.

Another kinetic instability, the Weibel instability~\citep{Weibel}, in contrast, is driven by temperature anisotropies, where the spread in the momentum of a species in one direction is higher than in another.
The difference in momentum spread can be characterized by bi-Maxwellian distributions as described in~\citep{Weibel} (often referred to as a thermal Weibel instability), or by the superposition of beams~\citep{Fried59} (also referred to as the current filamentation instability), associated with an effective temperature anisotropy. Although these two configurations are sometimes regarded as distinct instabilities, they can both be described, in the linear stage, by a single general dispersion relation.
The Weibel instability has been widely studied as a mechanism 
for spontaneous magnetic field generation in unmagnetized astrophysical environments and in the lab, in particular at collisionless shocks driven by counter-streaming plasmas~\citep{Medvedev1999, Spitkovsky2008, Spitkovsky2008b, Matsumoto2015, Ruyer2016, Fox2018, Pucci2021, Zhao2024}. 

The Weibel instability is also associated with magnetic reconnection~\citep{zenitani2008role,Swisdak2008,Liu2009}. 
The counter-streaming of reconnection jets through ambient plasma in pair plasmas generates a temperature anisotropy that drives Weibel-type electromagnetic modes, which modify current sheet dynamics.  Weibel-generated fields are also produced in laser-solid interactions used to study magnetic reconnection in the lab~\citep{Schoeffler2014,Schoeffler2016,Matteucci2018,Shukla2020}.

During the early studies of the tearing instability, the effect of temperature anisotropies was recognized~\citep{ForslundThesis}. Note that these are standard temperature anisotropies rather than the effective anisotropy explained in this paper. Ratios of the temperature perpendicular and parallel to the magnetic field greater than one ($T_\perp/T_\parallel > 1$) enhance the growth rate and lead to shorter-wavelength modes, while opposite anisotropies 
($T_\parallel/T_\perp > 1$) increase the wavelength and stabilize the tearing mode.
Studies confirmed this with magnetohydrodynamic (MHD) theory \citep{Shi1987} and kinetic simulations \citep{Ambrosiano1986}. This effect was further confirmed in
\citet{karimabadi2004role, Haijima2008,Quest2010,Matteini2013,Winarto2022}. 

We will show that these relations can be better understood by linking the Weibel instability, which naturally depends on this anisotropy (or an effective anisotropy), and the tearing instability. The associated drift of the current sheet leads to an effective temperature anisotropy (with higher temperature along the drift), analogous to the two-beam anisotropy inherent to the current filamentation instability, even when the standard temperature anisotropy is taken to be zero.

We now lay out the organization of the paper. After this introduction (Section~\ref{sec:intro}), we describe in Section~\ref{sec:setup} the Harris configuration and define some parameters that characterize the two instabilities. Next, in Section~\ref{sec:background}, we provide a theoretical background on the two instabilities.
We will then describe our simulation setup and computational parameters in Section~\ref{sec:sim-setup}. We explain our simulation results in Section~\ref{sec:results}. Finally, we will conclude with a discussion in Section~\ref{sec:conclusion}.

\section{Harris Equilibrium}\label{sec:setup}

We will consider in this paper the tearing instability starting from the Harris equilibrium.
Here, we describe the setup and define its key variables.

The equilibrium is composed of two regions with oppositely directed magnetic fields separated by a current sheet. The current sheet is composed of counter-drifting Maxwellian distributions of ions and electrons with uniform temperatures~$T_\alpha$ for the respective species~$\alpha$, drifting in opposite $\pm \hat{z}$-directions with a uniform speed~$v_{d\alpha}$.
The density profile (of both electrons and ions) in the Harris current sheet at $y = 0$ is given by
\begin{equation}
\label{Harrisn}
	n = n_0 {\rm sech}^2 \left(\frac{y}{a}\right) + n_b,
\end{equation}
where $n_0$ is the number density at the center of each current sheet for both ions and electrons. Since the diamagnetic drift determines the drift velocity $v_{d\alpha}$, the magnitude of $v_{d\alpha}/T_\alpha$ must be the same for each species. While we have included it in \eq{Harrisn} for generality, in this study, we will not consider any additional uniform background population, i.e., $n_b=0$.

The pressure of the current sheets is balanced by self-consistent magnetic field profiles, which results in a kinetic equilibrium. The magnetic field profile is given by
\begin{equation}
\label{HarrisB}
	B_x =B_0\tanh\left(\frac{y}{a}\right).
\end{equation}
where the asymptotic magnetic field balances the pressure at the center of the current sheet
\begin{equation}
\label{forcebalance}
    B_0 = \sqrt{8 \pi \sum_\alpha n_\alpha T_\alpha} = \sqrt{8 \pi n_0 T_\alpha\eta_\alpha}.
\end{equation}
Here, we sum over the respective electron and ion species, $\alpha = e$ and $i$. We then define $\eta_\alpha \equiv \left(\sum_{\alpha^\prime} T_{\alpha^\prime}\right)/T_\alpha$.
Using Amp\`ere's law, the current sheet half-thickness can be calculated to be
\begin{equation}
\label{ampere}
    a = \frac{c B_0}{4 \pi e \sum_\alpha n_\alpha v_{d\alpha}} = \frac{c B_0}{4 \pi e n_0 v_{d\alpha} \eta_\alpha}= \sqrt{\frac{T_\alpha c^2}{2 \pi n_0 e^2 v_{d\alpha}^2 \eta_\alpha  }},
\end{equation}
again, remembering that $v_{d\alpha}/T_\alpha$ is the same for each species, and assuming two species.

We can express the scales of the system as
the inertial length,
\begin{equation}
\label{classde}
    d_{\alpha} \equiv \frac{c}{\omega_{p\alpha}}=\sqrt{\frac{m_\alpha c^2}{4\pi n_0 e^2}},
\end{equation}
and the Larmor radius,
\begin{equation}
\label{classrho}
    \rho_{L\alpha} \equiv \frac{v_{T\alpha}}{\Omega_{c\alpha}} = \sqrt{\frac{1}{\eta_\alpha}} d_{\alpha} = \frac{v_{d\alpha}}{v_{T\alpha}}a = M_{\alpha z}a,
\end{equation}
where, aside from the definition, \eq{classrho} is specific to the Harris equilibrium.
Here $\omega_{p\alpha}$ is the plasma frequency of species $\alpha$,
$\Omega_{c\alpha} = e B_0/(m_\alpha c)$ is the cyclotron frequency, $v_{T\alpha}/c \equiv \sqrt{2T_\alpha/(m_\alpha c^2)}$ is the thermal velocity, and we define $M_{\alpha z} \equiv v_{d\alpha}/v_{T\alpha}$ as the thermal Mach number of the drift, which will be useful when comparing with the Weibel instability.

We note that our setup here has antiparallel magnetic fields. Often, one would include a uniform perpendicular magnetic field along the $z$ direction, known as a guide field. For simplicity, we will assume that this is zero, leaving the effects of a guide field for future work.

\section{Theoretical Background}\label{sec:background}

\subsection{Tearing instability}
The general, non-relativistic collisionless tearing instability, for a Harris equilibrium with no guide field from \cite{Zelenyi1979} is equivalent to predictions from \cite{Laval1966} and \cite{Coppi1966}, except for a different $k$ dependence and a numerical factor. This growth rate is given by the following expression
\begin{equation}
\label{tearingrho}
\frac{\gamma a}{v_{Te}} = \left(\frac{1}{\pi}\right)^{1/2} f(k) \left(\frac{\rho_{Le}}{a}\right)^{3/2} \eta_e\frac{1 }{1 + (\mu T_e/T_i)^{1/4}} 
\end{equation}
where
\begin{equation}
\eta_e = 1 + T_i/T_e,
\end{equation}
\begin{equation}
\mu = m_e/m_i,
\end{equation}
and
\begin{equation}
f(k) = ka(1-k^2a^2)
\end{equation}
thus the growth reaches a peak at
\begin{equation}
ka = \frac{1}{\sqrt{3}} 
\end{equation}
This theoretical prediction was verified with particle-in-cell simulations for pair plasmas ($\mu =1$, $\eta_e = 2$) in~\citet{Schoeffler2025} and is further verified in the present paper.

One should keep in mind that this prediction of the growth rate assumes the constant-$\psi$ approximation \citep{Burkhart1989}, which is valid in the limit that $ka \approx 1$. It is applicable down to values of $ka\sim k_{max}a$ corresponding to the maximum growth rate, below which the instability develops in the large $\Delta'$ regime (see e.g.~\citealt{DelSarto2016} and references therein). Therefore, as also described in~\citet{Schoeffler2025}, the prediction that $ka \approx 1/\sqrt{3}$ is only an estimate. Several analytical and numerical studies have more accurately predicted a dispersion relation \citep{Chen1985,Daughton1999,Daughton2003,Daughton2005,Petri2007}, finding the wave number remains close to $k_{max}a \sim 1/2$. In addition, this is consistent with the idea that the simulation-predicted wave-vector is at the transition between the constant-$\psi$ and regime of the maximum growth rate (see Fig.~4 \citealt{2016JPlPh..82e5301T} for the resistive tearing case, and Fig.~1 \citealt{DelSarto2016} for the collisionless case).

Equivalently, the tearing dispersion relation can be written in terms of the electron Mach number by the relation $M_{ez}=\rho_{Le}/a$ from \eq{classrho},
\begin{equation}
\label{tearingkMez}
\frac{\gamma(k)a}{v_{Te}} = \frac{\gamma d_e}{v_{Te}M_{ez}\eta_e^{1/2}} = \left(\frac{1}{\pi}\right)^{1/2} f(k) M_{ez}^{3/2} \eta_e \frac{1 }{1 + (\mu T_e/T_i)^{1/4}} 
\end{equation}
where
\begin{equation}
f(k) = kd_e(M_{ez}^2\eta_e)^{-3/2}(M_{ez}^2\eta_e-k^2d_e^2)
\end{equation}
with a peak growth at
\begin{equation}
\label{tearingk}
kd_e = \sqrt{\frac{M_{ez}^2\eta_e}{3}}.
\end{equation}
One can then write the peak normalized growth rate 
as
\begin{equation}
\label{tearingMez}
\frac{\gamma a}{v_{Te}} = \left(\frac{1}{\pi}\right)^{1/2}\frac{2}{3^{3/2}} M_{ez}^{3/2} \eta_e \frac{1 }{1 + (\mu T_e/T_i)^{1/4}}
\end{equation}

\subsection{Weibel instability}\label{sec:weibel}
The Weibel instability is often put into two categories,
the thermal Weibel instability~\citep{Weibel}, where the initial setup is bi-Maxwellian with a so-called hot direction and a cold direction,
and the current filamentation instability~\citep{Fried59}, where the hot direction is made hot by a drift between two species modifying the total thermal spread in one direction.

Let us consider a general set of bi-Maxwellians of species $\alpha$, with a temperature along the $z$ direction $T_{\alpha z}$, a perpendicular temperature $T_{\alpha \perp}$, and a drift $v_{d\alpha }$ along $z$. This setup will thus include both types of instabilities. The dispersion along the electromagnetic branch can be written as
\begin{equation}
k^2c^2 -\omega^2 + \sum_\alpha \omega_{p\alpha}^2 + \sum_\alpha \omega_{p\alpha}^2 \left(A_\alpha + 1 + 2M_{\alpha z}^2\right) W\left(\xi_\alpha\right)=0
\label{full-disp}
\end{equation}
where $A_\alpha = T_{\alpha z}/T_{\alpha \perp}-1$ is the anisotropy of species $\alpha$,
\begin{equation} \xi_\alpha =
        \frac{\omega}{kv_{T\alpha\perp}} \text{,}
\end{equation}
\begin{equation} M_{\alpha z} =
        \frac{v_{d\alpha}}{v_{T\alpha\perp}}  \text{,}
\end{equation}
and
\begin{equation}
\label{Wdefined}
        W\left(\xi\right) \equiv -1 -\xi Z\left(\xi\right) = \frac{1}{2} Z^\prime\left(\xi\right)
\end{equation}
is a convenient function written in terms of the plasma dispersion function $Z(\xi)$.

One should note that in Weibel's original paper~\citep{Weibel}, he only considers one species, i.e. electrons. People normally refer to an unstable mode dominated by one species as the Weibel instability, i.e., the electron or ion Weibel instability. Furthermore, he does not consider a drift between species ($M_{\alpha z}=0$), and therefore does not define a different temperature along the drift, which we take in this paper as the hot direction. Because it does not complicate the calculation too much, he includes a magnetic field defining the parallel direction, which is considered cold, and the other two are considered hot. 

In Weibel's definition, \eq{full-disp} remains the same, but because of the magnetic field, he redefines
\begin{equation}
\label{altxi}
\xi_{\alpha,\rm{Weibel}} = \frac{\omega \pm \Omega_{c\alpha}}{kv_{T\alpha\parallel,\rm{Weibel}}},
\end{equation}
and $A_{\alpha,\rm{Weibel}} = T_{\alpha \perp,\rm{Weibel}}/T_{\alpha \parallel,\rm{Weibel}}-1$,
with the addition of a final term
\begin{equation}
\label{magnetterm}
+ \sum_\alpha \omega_{p\alpha}^2 \xi_{\alpha 0} Z\left(\xi_\alpha\right).
\end{equation}
where
\begin{equation}
\xi_{\alpha0} = \frac{\pm \Omega_{c\alpha}}{kv_{T\alpha\parallel,\rm{Weibel}}}.
\end{equation}

This relation will prove useful later, because in our setup, a magnetic field is present everywhere except at the center of the current sheet. When one includes both electron and ion species and a magnetic field, but no drifts, the dispersion relation is divided into two modes: a right-hand-circularly-polarized (RHCP) mode that corresponds to the Whistler instability, and a left-hand-circularly-polarized (LHCP) mode that corresponds to the ion cyclotron instability. For RHCP, the sign of $\xi_{e0}$ is negative and $\xi_{i0}$ is positive, while for LHCP, the signs are switched. 

Unmagnetized, the physics of the instability remains unchanged if there are two hot directions or two cold directions; the unstable modes consist of a growing magnetic field perpendicular to both $k$ (along one of the cold directions) and to one of the hot directions. We will consider the mode where the hot direction is along the current $\hat{z}$, and the cold direction is along the direction of the equilibrium magnetic field $\hat{x}$. The temperature in the $\hat{y}$ direction does not play an important role for us. 

The Weibel instability in \eq{full-disp}, includes the driving term of the thermal Weibel instability, $A_\alpha$, and for the current filamentation instability, $2M_{\alpha z}^2$. The two instabilities have identical dispersion relations, where an effective anisotropy can be defined as
\begin{equation}
\label{Aeff}
A_{\alpha ,\rm{eff}} \equiv \frac{T_{\alpha z,\rm{eff}}}{T_{\alpha \perp}} - 1 = 2M_{\alpha z}^2.  
\end{equation}

Properly including a drift becomes more challenging when a magnetic field is included, but we will nevertheless use the theoretical growth rate with this effective anisotropy for comparisons with our simulation results. 

In our setup, we will use our definition of $A_\alpha$ (rather than Weibel's) and replace $v_{T\alpha\parallel,\rm{Weibel}}$ in \eqs{altxi}{magnetterm} with $v_{T\alpha\perp}$, thus retaining it as the "cold" direction.
Furthermore, we do not include this standard anisotropy, but only the effective one from \eq{Aeff}. 

In the appendices, we derive the limits of the Weibel instability in the absence of a magnetic field for the electron-only case (ions are considered immobile) and for both electron and ion motion for different limits of $\xi_{\alpha}$.
In particular, for $\xi_{\alpha} \ll 1$, we derive \eq{lowxigrowthionfmodeApp} or
\begin{equation}
\label{lowxigrowthionfmode}
\frac{\gamma a}{v_{Te}} = 2 \left(\frac{2}{3}\right)^{3/2}
\left(\frac{1}{\pi}\right)^{1/2}
\frac{1}{1+\left(\mu T_e/T_i\right)^{1/2}} 
\eta_e
M_{ez}^2,
\end{equation}
comparable with \eq{tearingMez}
and
\begin{equation}
\label{lowxikmaxion}
kd_e = \sqrt{\frac{2M_{ez}^2\eta_e}{3},} 
\end{equation}
comparable with \eq{tearingk}.

\subsection{tearing/Weibel comparison}
We now compare the Weibel instability with the tearing instability.
As we saw in \eq{Aeff}, the anisotropy that drives the Weibel instability is $A_{\alpha,\rm{eff}} \sim M_{\alpha z}^2 = v_{d \alpha}^2/v_{T \alpha}^2$ and therefore we also expect the Weibel instability to occur in the tearing setup, due to the drift associated with the current.
In the regime where the current sheet is thick, $a/\rho_{Li} \gg 1$, this is equivalent to saying that $M_{iz} \ll 1$, since $\rho_{Li}/a = M_{iz}$ from \eq{classrho}, and consequentially $\rho_{Le}/a = M_{ez} \ll 1$ as well.

The wavenumber of the Weibel instability is directed along the so-called cold direction. The growing mode is composed of a magnetic field perpendicular to both this wavenumber and the hot direction.
This is all consistent with the tearing mode, where the hot direction is along the current drift ($z$ direction), and the wavenumber is along the current sheet perpendicular to the drift ($x$ direction). The mode is composed of a magnetic field perpendicular to both of these ($y$ direction).

The form of the equations for the Weibel growth rate \eq{lowxigrowthionfmode} and the tearing growth rate \eq{tearingMez} is quite similar in the regime $a/\rho_{Li},~a/\rho_{Le} \gg 1$,
but the scaling is not the same. The tearing instability scales as $\gamma a/v_{Te} \sim M_{ez}^{3/2}$, while the Weibel instability scales as
$\sim M_{ez}^{2}$. Therefore, for the thicker current sheets, the Weibel instability is slower by a factor $\sim M_{ez}^{1/2}$, and the tearing mode should dominate.
The ratio of the Weibel and tearing growth rates is
\begin{equation}
\label{lowxigrowthratio}
\frac{\gamma_W}{\gamma_T} = 
2^{3/2}M_{ez}^{1/2}\frac{1+\left(\mu T_e/T_i\right)^{1/4}}{1+\left(\mu T_e/T_i\right)^{1/2}} 
\end{equation}
In addition to the dependence on the drift velocity, there is also a dependence on the temperature ratio. As long as the ratio $T_e/T_i < 1/\mu$, which is true in most cases, this dependence is negligible. However, for very hot electron temperatures, the Weibel instability is a factor $(\mu T_e/T_i)^{-1/4}$ slower. More significantly, when $M_{ez}$ becomes of order unity, the growth rates of the two instabilities become of the same order, and the Weibel physics should play a significant role.

In the very thin current sheet regime ($a/\rho_{Le} < 1$, or equivalently $M_{ez} \gg 1$, associated with the non-resonant mode $\xi_e \gg 1$), the growth rate of the Weibel instability scales as $\gamma a/v_{Te} \sim 1$, i.e. independent of $a/d_e$ [See \eq{highxigrowthion}]. For more intermediate current thickness $a/\sqrt{\rho_{Le}\rho_{Li}} < 1$, there is a similar independence for the resonant mode ($\xi_e \ll 1$) [See \eq{lowxigrowthionfmodesmalla}.]. Similarly, we will show that the measured instability reaches a maximum growth rate here; $\gamma a/v_{Te}$ loses its dependence on $M_{ez}^{3/2}$ and becomes constant. More accurately, we will show that it follows the prediction of the magnetized LHCP (ion cyclotron branch) [\eqs{full-disp}{magnetterm}].

Another important comparison to make is the predicted wavelength.
For thick current sheets (small $A_{\rm{eff}}$), the wavenumber for tearing should be around $ka \approx 0.5$. Likewise, for Weibel, $kd_e \approx \sqrt{A_{\rm{eff}}/3} \sim M_{ez}$ (or $ka \sim 1$), so
the wavenumbers of the two modes scale similarly.
For Weibel in the high $A_{\rm{eff}}$ regime, corresponding to thin current sheets, $kd_e \sim A_{\rm{eff}}^{1/4}$, and thus we expect
$ka \sim M_{ez}^{-1/2} \sim (a/\rho_{Le})^{1/2}$ [See \eq{highxikmaxion}.] which in principle corresponds to a wavenumber smaller than $ka \sim 1$. The wavenumber of tearing in the thin current sheet may also be smaller than
$ka \sim 1$, which can be tested.

Although the two instabilities are not identical, they are certainly interconnected. This naturally explains why an anisotropic current sheet will have an effect on the growth rate of the tearing instability. When the temperature is higher along the current direction, the instability grows faster, and the wavelength is smaller ($ka$ is larger).  When the anisotropy is reversed, and it is hotter along the $k$ direction, the instability slows down and is eventually suppressed, while the fastest growing wavelength increases. 

In our discussions so far, we have neglected the effects of the magnetic field. The bounce frequency of a particle of species $\alpha$ moving along the $y$ direction, assuming $v_z$ is held constant at $v_{d\alpha}$ (This is reasonable if $v_{d\alpha} \gg v_{T\alpha}$, as seen in thin current sheets.), is given by $\omega_B = v_{T\alpha}/a$. The amplitude of such bounces is $\sim a$, so the particles will experience magnetic fields of order $B_0$. Since the growth rate is less than the bounce frequency of an electron ($\gamma/\omega_B = \gamma a/v_{Te} < 1$), for all cases considered, one should also take into account the effect of the magnetic field to get a more accurate estimate of the growth rate.

In the opposite thick-current-sheet regime ($v_{d\alpha} \ll v_{T\alpha}$), the bounce frequency is given by Speiser orbits with a frequency
\begin{equation}
\omega_B = \sqrt{\frac{\Omega_{c\alpha}v_{T\alpha}}{a}} = \frac{v_{T\alpha}}{a} \sqrt{\frac{a}{\rho_{L\alpha}}}
\end{equation}
which is even faster. However, the amplitude of the bounce is $\sim \sqrt{a \rho_{L\alpha}}$, and therefore particles experience relatively weak fields ($B \ll B_0$) in this regime.

\section{Simulation Setup}\label{sec:sim-setup}
We employ the semi-implicit fully kinetic particle-in-cell code
ECsim \citep{Lapenta2017, Boella2026} to investigate the kinetic evolution of current sheets
across a range of thicknesses spanning from ion to sub-electron
scales. The simulations are performed in a 2D configuration
with periodic boundary conditions in all directions, similar to \cite{Schoeffler2025,Schoeffler2025b}.
The initial configuration consists of a double Harris current
sheet; we thus modify \eq{Harrisn} and \eq{HarrisB} to
\begin{equation}
    B_x(y) = B_0 \left[ \tanh\left(\frac{y - y_1}{a}\right)
    - \tanh\left(\frac{y - y_2}{a}\right) - 1 \right],
\end{equation}
and
\begin{equation}
    n(y) = n_0 \left[ \mathrm{sech}^2\left(\frac{y - y_1}{a}
    \right) + \mathrm{sech}^2\left(\frac{y - y_2}{a}\right)
    \right],
\end{equation}
where the two current sheets are
centered at $y_1 = L_y/4$ and $y_2 = 3L_y/4$. Both
species are initialized with equal temperatures $T_e = T_i$, corresponding to thermal velocities $v_{Te} = 0.035\,c$ and $v_{Ti} = v_{Te}\sqrt{m_e/m_i}$.

The simulation domain is scaled with the current sheet thickness, with $L_x/a = 50$ and $L_y/a = 100$, ensuring that the same range of normalized wavenumbers~$ka$ is resolved across all runs.
We perform simulations for four mass ratios:
$m_i/m_e = 25$, $256$, $512$, and~$1836$. For
$m_i/m_e = 25$ and~$256$, the current sheet half-thickness is
varied over $a/\rho_{Li} = 1.0$, $0.9$, $0.7$, $0.5$, $0.3$, $0.1$,
and~$0.05$. For $m_i/m_e = 512$ and~$1836$, the half-thickness is varied over a reduced set of $a/\rho_{Li} = 1.0$, $0.3$, $0.1$, and~$0.05$.

The grid resolution is $\Delta x/a = \Delta y/a = 0.25$ with $N_x \times N_y = 200 \times 400$ cells, and the time step is $\Delta t \, c/a = 0.1$. Each species is initialized with 1024 particles per cell.  The time step satisfies
$v_{Te} \, \Delta t / \Delta x = 0.014$, ensuring that thermal particles traverse a small fraction of a cell per step.

In addition to the thickness scan, we perform two additional thick-sheet runs at
$a/\rho_{Li} = 2.5$, for $m_i/m_e = 25$ and~$256$. The parameters of these two runs do not follow the setup above: the domain is smaller 
($L_x/a =40,\ L_y/a = 80$), with higher
resolution ($N_x \times N_y = 384 \times 768$ cells, giving
$\Delta x/a = \Delta y/a \approx 0.10$). The
time step is $\Delta t\,c/a = 0.04$, with $1024$ particles per cell per
species. The time step satisfies
$v_{Te} \, \Delta t / \Delta x = 0.013$, ensuring that thermal particles still traverse a small fraction of a cell per step.
\section{Simulation results}\label{sec:results}
We present results from two-dimensional particle-in-cell
simulations of Harris current sheets across the parameter
range described in Section~\ref{sec:sim-setup}. Before comparing the measured growth rates with
theoretical predictions across all current sheet thicknesses
and mass ratios, we first examine the energy evolution and spatial structure for two representative cases at
$m_i/m_e = 256$.
Figure~\ref{fig:energy_evolution} shows the temporal evolution
of energy components for current sheets with $a/\rho_{Li} = 1.0$
and $a/\rho_{Li} = 0.1$ at mass ratio $m_i/m_e = 256$. The top
panels display the magnetic ($E_B$), electric ($E_E$), kinetic
($E_{KE}$), and total ($E_{tot}$) energies normalized by the
initial total energy, which is conserved to machine precision. The bottom panels show the absolute
change in each component on a logarithmic scale. To isolate the tearing mode, we also display the magnetic ($E_{By}$) energy solely due to the $B_y$ component of the field. Furthermore, we isolate unstable long-wavelength modes from short-wavelength noise in $E_{By}$ by performing a one-dimensional Fourier transform along~$x$ and retaining only modes with $k_x a < 1$, as done in \cite{Schoeffler2025,Schoeffler2025b}.

For the $a/\rho_{Li} = 1.0$ case (panels~a and~c), the system is
initially magnetically dominated, with
$E_B/E_{tot}(0) \approx 0.93$, as there is no background population. The filtered $B_y$ energy
exhibits clear exponential growth in three stages. In the linear stage, from
$t v_{Te}/a = 254$ to $507$
($t\gamma_T \approx [1$, $2$]), an exponential fit yields
$\gamma_{\mathrm{lin}}\,a/v_{Te} \approx 3.53\times10^{-3}$, i.e.
$\gamma_{\mathrm{lin}}/\gamma_T \approx 1.09$, where $\gamma_T$ is the theoretical tearing growth rate  from Eq.~(\ref{tearingrho}). The growth then
slows in an intermediate stage between $t v_{Te}/a = 634$ and $1015$ ($t\gamma_T =$ [$2.5$, $4$]), with
$\gamma_{\mathrm{int}}\,a/v_{Te} \approx 1.84\times10^{-3}$
($\gamma_{\mathrm{int}}/\gamma_{\rm lin} \approx 0.52$). Finally, it reaches a fast
nonlinear stage between $t v_{Te}/a = 1243$ and $1319$,
with
$\gamma_{\mathrm{nl}}\,a/v_{Te} \approx 1.22\times10^{-2}$
($\gamma_{\mathrm{nl}}/\gamma_{\rm lin} \approx 3.5$), and then it saturates near
$tv_{Te}/a \approx 1523$.

The $a/\rho_{Li} = 0.1$ case (panels b and d) differs considerably. A significant portion of the initial energy is in the particles, with
$E_{KE}/E_{\mathrm{tot}}(0) \approx 0.59$ and
$E_B/E_{\mathrm{tot}}(0) \approx 0.41$. This is because the Harris equilibrium fixes a current-sheet drift inversely proportional to the thickness [See \eq{classrho}], so for thin enough sheets the (ion-dominated) bulk-flow kinetic energy becomes comparable to the magnetic energy.
The filtered $B_y$ energy still grows
exponentially, from 
$t v_{Te}/a = 80.2$ to $281$
($t \gamma_{\rm IC} = [2.2, 7.9]$),
but at a rate
$\gamma_{\mathrm{lin}}\,a/v_{Te} \approx 2.24\times10^{-2}$
($\gamma_{\mathrm{lin}}/\gamma_T \approx 0.22$), well below the tearing
prediction, which is understandable as this prediction assumes $a/\rho_{Li} \gg 1$. However,
$\gamma_{\mathrm{lin}}/\gamma_{\rm IC} \approx 0.80$, where $\gamma_{\rm IC}$ is the
LHCP ion cyclotron growth rate given by \eqs{full-disp}{magnetterm}, which gives a
good estimate of the growth rate. The growth then slows to
$\gamma_{\mathrm{int}}\,a/v_{Te} \approx 1.0\times10^{-2}$
($\gamma_{\mathrm{int}}/\gamma_{\rm lin} \approx 0.45$) between
$t v_{Te}/a = 281$ and $481$,
and then it saturates near 
$t v_{Te}/a \approx 521$
without reaching a fast nonlinear stage.

We identify these three stages and measure the respective growth rates in all simulations. When there were only two stages, if the second is faster, it is considered the third fast nonlinear stage, and if it is slower (like our example with $a/\rho_{Li} = 0.1$, $m_i/m_e = 256$), it is considered the second slow nonlinear stage. For two cases (when $a/\rho_{Li} = 0.05$, for $m_i/m_e = 25$ and $256$), only the linear stage exists. {Table~\ref{tab:three_phase_growth_rates} shows the measured
growth rates across all mass ratios and current sheet
thicknesses.

{Figure~\ref{fig:By_saturation} presents the spatial structure of the magnetic field component $B_y$ during the linear stage for both $a/\rho_{Li} = 1.0$ and $a/\rho_{Li} = 0.1$ with $m_i/m_e=256$.  
Panel (a) shows the magnetic field structure for the $a/\rho_{Li} = 1$ case. The $B_y$ perturbations exhibit well-defined, coherent structures characteristic of classical tearing mode islands.}
The dominant wavelength along the $x$-direction ($ka \sim 0.6$) is consistent with the fastest-growing tearing mode predicted by linear theory.
Panel (b) displays the magnetic structure for the $a/\rho_{Li} = 0.1$ case, which is qualitatively the same, except for a stronger, uniformly distributed noise. The wave number ($ka \sim 0.5$) is still consistent with theory, but is slightly less, also consistent with the predicted trend for $a/\rho_{Li} < 1$
where for thin enough current sheets $ka \sim M_{ez}^{-1/2}$ [See \eq{highxikmaxion}.].

\begin{figure}
    \centering
    \includegraphics[width=\textwidth]{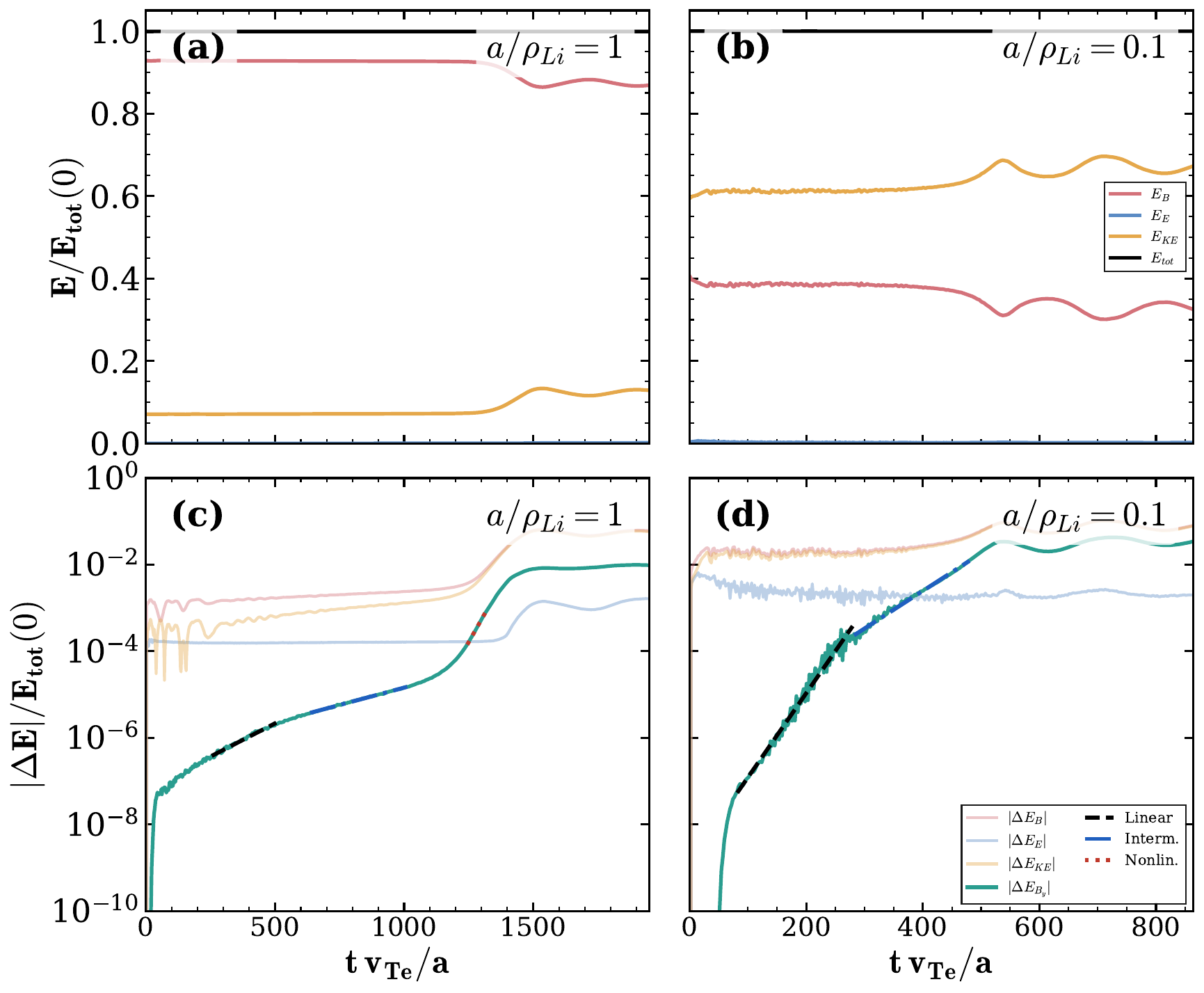}
    \caption{Energy evolution for current sheets with (a,c) $a/\rho_{Li} = 1$ and (b,d) $a/\rho_{Li} = 0.1$ at mass ratio $m_i/m_e = 256$. Top row: temporal evolution of magnetic ($E_B$, red), electric ($E_E$, blue), kinetic ($E_{KE}$, yellow), and total ($E_{tot}$, black) energy normalized by the initial total energy. Bottom row: absolute change in energy components on a logarithmic scale. The  $B_y$ energy is shown in green, with the dashed black line indicating the exponential fit during the linear growth phase, blue indicating the intermediate phase, and red indicating the fast-growing non-linear phase.}
    \label{fig:energy_evolution}
\end{figure}

\begin{table}
  \centering
  \label{tab:three_phase_growth_rates}
  \begin{tabular}{c c c c c c c}
    \hline\hline
    $m_i/m_e$ & $a/\rho_{Li}$ & $\gamma_{\rm lin}a/v_{Te}$ & $\gamma_{\rm int}a/v_{Te}$ & $\gamma_{\rm nl}a/v_{Te}$ & $\gamma_{T}a/v_{Te}$ & $\gamma_{\rm IC}a/v_{Te}$\\
    \hline
    \multirow{7}{*}{25}
       & 1.0  & 0.0228 & 0.0156 & 0.0524 & 0.0160 & 0.00869\\
       & 0.9  & 0.0284 & 0.0186 & 0.0576 & 0.0187 & 0.0125\\
       & 0.7  & 0.0308 & 0.0201 & 0.0513 & 0.0273 & 0.0231\\
       & 0.5  & 0.0501 & 0.0156 & 0.0608 & 0.0451 & 0.0376\\
       & 0.3  & 0.0603 & 0.0425 & 0.0454 & 0.0971 & 0.0558\\
       & 0.1  & 0.0523 & --- & 0.0547 & 0.505 & 0.0605\\
       & 0.05 & 0.0193 & 0.0103 & --- & 1.43 & 0.0421\\
      \hline
    \multirow{7}{*}{256}
       & 1.0  & 0.00353 & 0.00184 & 0.0122 & 0.00323 & 0.00270\\
       & 0.9  & 0.00475 & 0.00264 & 0.0129 & 0.00378 & 0.00396\\
       & 0.7  & 0.00796 & 0.00372 & 0.0122 & 0.00551 & 0.00738\\
       & 0.5  & 0.0112 & 0.00528 & 0.0154 & 0.00913 & 0.012\\
       & 0.3  & 0.0176 & 0.00819 & 0.0172 & 0.0196 & 0.0187\\
       & 0.1  & 0.0224 & 0.01 & --- & 0.102 & 0.0279\\
       & 0.05 & 0.0191 & --- & --- & 0.289 & 0.0266\\
      \hline
    \multirow{4}{*}{512}
       & 1.0  & 0.00165 & 0.000895 & 0.00773 & 0.00198 & 0.00194\\
       & 0.3  & 0.00964 & 0.00533 & 0.0107 & 0.0121 & 0.0133\\
       & 0.1  & 0.0179 & 0.00622 & 0.0118 & 0.0627 & 0.0204\\
       & 0.05 & 0.0143 & 0.00675 & --- & 0.177 & 0.0212\\
      \hline
    \multirow{4}{*}{1836}
       & 1.0  & 0.000307 & --- & 0.00366 & 0.000799 & 0.00103\\
       & 0.3  & 0.00285 & --- & 0.00512 & 0.00486 & 0.00705\\
       & 0.1  & 0.00795 & 0.00368 & 0.00677 & 0.0252 & 0.0112\\
       & 0.05 & 0.0075 & 0.00359 & 0.0061 & 0.0714 & 0.0125\\
      \hline
    \hline
  \end{tabular}
  \caption{Normalized growth rates $\gamma\,a/v_{Te}$ from filtered $B_y$ ($ka<1$) for the linear, intermediate, and nonlinear phases, as well as the theoretical values for the tearing [\eq{tearingrho}] (using $a/\rho_{Le} = a/\rho_{Li} \sqrt{T_i/T_e}\sqrt{m_i/m_e}$ [\eq{classrho}], $T_i/T_e = 1$, and $\eta_e = 2$) and ion cyclotron [\eqs{full-disp}{magnetterm}] (using $\xi_{e} = \xi_i \sqrt{T_i/T_e}\sqrt{m_e/m_i}$, $\xi_{0\alpha} = 1/k\rho_{L\alpha}$, $A_\alpha=0$, $M_\alpha^2 = \rho_{L\alpha}^2/a^2$, $\omega_{p\alpha}^2a^2/c^2 = a^2/\eta_\alpha \rho_{L\alpha}^2$ [\eq{classrho}], $\eta_i = \eta_e = 2$,  and $v_{Te}/c = 0.035$) instabilities. }
\end{table}

\begin{figure}
    \centering
    \includegraphics[width=\textwidth]{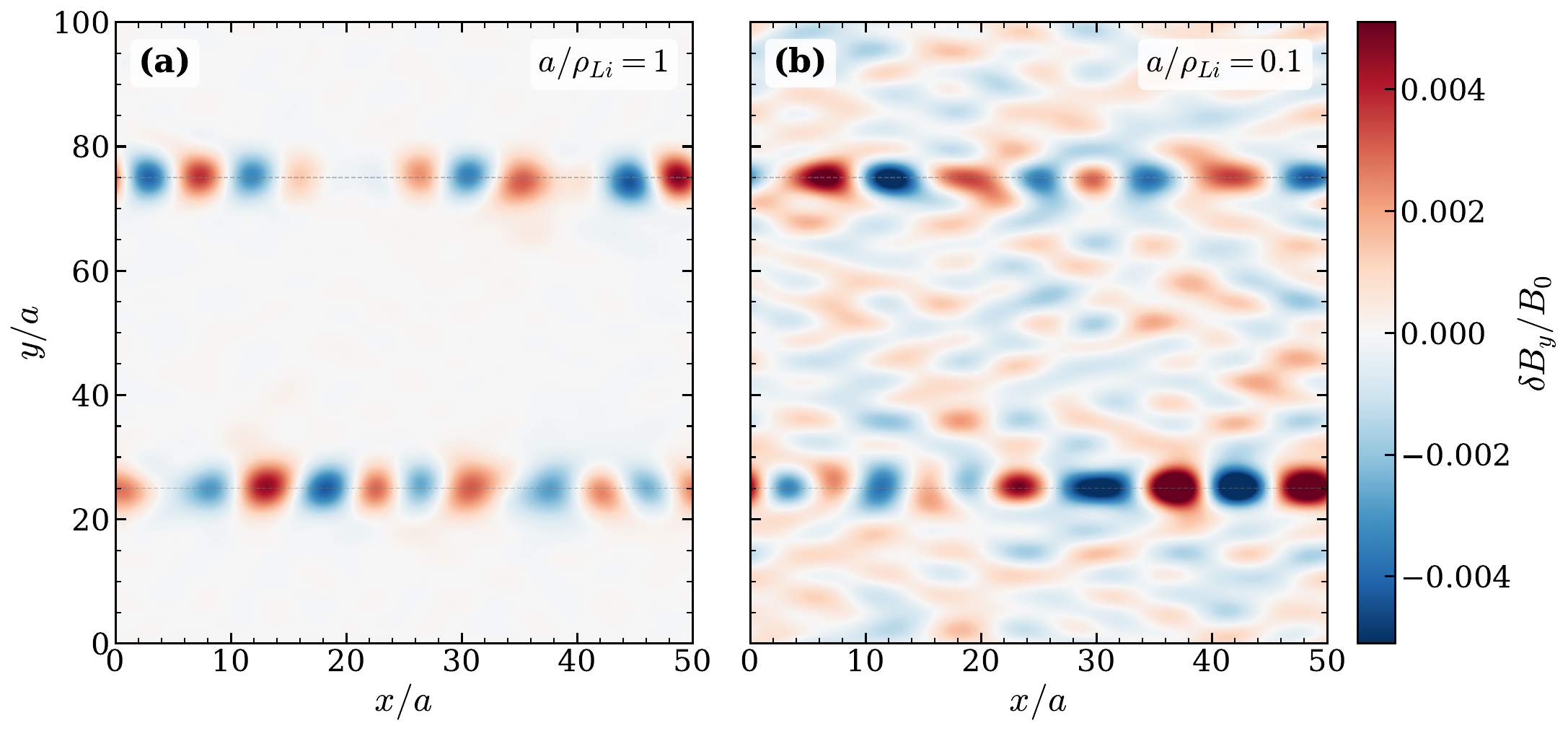}
    \caption{Spatial structure of the normalized perturbation magnetic field component $\delta B_y/B_0$ during the linear growth phase for (a) $a/\rho_{Li} = 1$ at $t v_{Te}/a = 240$ and (b) $a/\rho_{Li} = 0.1$ at $tv_{Te}/a = 107$.} 
    \label{fig:By_saturation}
\end{figure}

Figure~\ref{fig:growth_rate_CS_thickness} compares normalized linear growth rates $\gamma_{\rm lin} a / v_{Te}$ from PIC simulations (yellow circles) with theoretical predictions for the tearing (blue) and Weibel instabilities across the full scan in current sheet half-thickness $a/\rho_{Li}$, for each mass ratio $m_i/m_e = 25$, $256$, $512$, and $1836$.
The curves include the limits on the electron--ion Weibel instability for both resonant ($\xi \ll 1$) [\eq{lowxigrowthion}] plotted in dotted purple (with the large $a/d_e$ limit in dashed black) and non-resonant ($\xi \gg 1$)  [\eq{highxigrowthion}] plotted in dotted black, the full numerical dispersion solution
for a homogeneous, unmagnetized plasma (black) [\eq{full-disp}], and the magnetized
whistler/RHCP (orange) and ion cyclotron/LHCP (red) branches, using the magnetized definition of $\xi_\alpha$ [\eq{altxi}] and including the term in \eq{magnetterm}, but using the original definitions of directions defined as parallel to the drift velocity associated with the current (we assume $A_\alpha=0$ or $v_{T\alpha\parallel} = v_{T\alpha\perp}$, and use the effective anisotropy given by $2M_\alpha^2$).
Remember that $M_i^2 = \rho_{Li}^2/a^2$
and $M_e^2 = M_i^2 (T_e/T_i)(m_e/m_i)$ from \eq{classrho}.

For thick current sheets ($a/\rho_{Li} \gg 1$), the theoretical tearing growth rate in blue [Figure~\ref{fig:growth_rate_CS_thickness}] is much greater than the Weibel growth rate shown in solid black, which in this regime is well approximated by the estimation ($\xi \gg 1$) in dashed black. The two growth rates become similar when $a/\rho_{Li} \sim 1$. 
In this moderately thin current regime, the theoretical ion-cyclotron branch growth rate in red also has a similar magnitude. The rate is maximized here, limited to a peak growth close to the resonant ($\xi \ll 1$) limit of the Weibel instability [limit of \eq{lowxigrowthion}, given by \eq{lowxigrowthionfmodesmalla} shown in dotted purple]. For very thin current sheets, the theoretical Weibel growth rate matches the whistler branch shown in orange.

The measured linear growth rate is close to the tearing prediction in blue, for thick current sheets ($a/\rho_{Li} \geq 0.5$),
($\gamma_{\rm lin}/\gamma_T \approx 1$). We also confirm this with our additional thick-current-sheet simulation with $a/\rho_{Li} = 2.5$, where the tearing growth rate is much faster than the Weibel instability.
When $a/\rho_{Li} \sim 1$, the measured linear growth rate is similar to both the Weibel and tearing predictions.
However, as the current sheet thickness decreases below $a/\rho_{Li}=0.5$, the linear growth rate loses the $(a/\rho_{Li})^{-3/2}$ dependence from the tearing prediction (which assumes $a/\rho_{Li}\gg 1$), and also does not match the $(a/\rho_{Li})^{-2}$ scaling predicted by the Weibel instability. 

The growth rate does, however, match the prediction for the ion cyclotron/LHCP branch (within a factor of 2) [\eqs{full-disp}{magnetterm}]. This agreement is consistent across all mass ratios and indicates that the finite $B_x$ surrounding the current sheet center plays a significant role in setting the growth rate in the thin-sheet regime, rather than the unmagnetized Weibel prediction alone.
The low $\xi$ approximation of the unmagnetized ion-electron Weibel growth rate [\eq{lowxigrowthion}] gives a reasonable estimate for the peak growth rate as a function of $a/\rho_{Li}$, in the asymptotic regime, which is predicted in \eq{lowxigrowthionfmodesmalla}.
For realistic mass ratios, the measured growth rate appears to match the Weibel predictions better, but it is not clear if this is merely coincidence, as it also remains within a factor of 2 of the tearing/ion cyclotron (a more detailed convergence test over resolutions is needed to distinguish the two).

The red circles in Figure~\ref{fig:growth_rate_CS_thickness} correspond to the measurements of the nonlinear growth rates ($\gamma_{\rm nl} a/v_{Te}$), when available, which seem to be weakly dependent on $a/\rho_{Li}$. These nonlinear growth rates
follow the trend reported in \cite{Schoeffler2025} and are likely explained by the mechanism proposed therein; as the unstable "tearing" modes reach a nonlinear stage, the magnetic field (and density) in the center of the x-point decreases. The local value of $a/\rho_{Li}$ (and $a/d_{i}$) thus decreases until it reaches a maximum growth rate (close to the peak linear growth rate as a function of $a/\rho_{Li}$, as seen in the yellow circles).
A nonlinear growth rate only exists for thick current sheets where the linear (or intermediate) growth rate is smaller than the nonlinear one. The nonlinear growth rate approaches the linear one at current thicknesses around $a < \sqrt{\rho_{Le} \rho_{Li}}$.

\begin{figure}
    \centering    \includegraphics[width=0.8\textwidth]{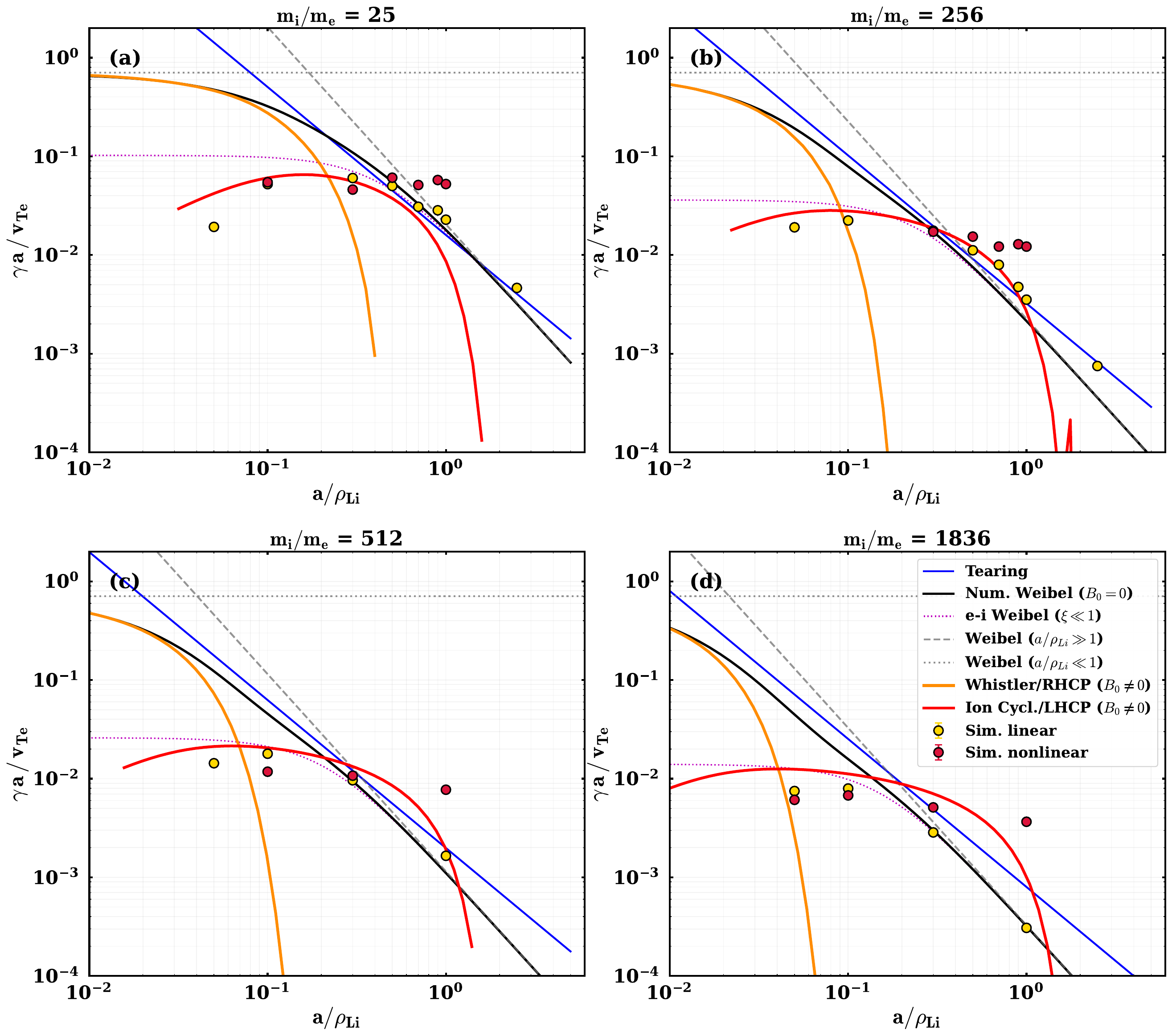}
    \caption{Normalized growth rate $\gamma a / v_{Te}$ as a function of the current sheet half-thickness $a/\rho_{Li}$ for mass ratios (a) $m_i/m_e = 25$, (b) $m_i/m_e = 256$, (c) $m_i/m_e = 512$, and (d) $m_i/m_e = 1836$. 
    The theory for tearing with a thick current sheet [\eq{tearingrho}] is plotted in blue. The black curve is the numerical solution of the Weibel instability from \eq{full-disp}.
    The limits of small $\xi$ [\eq{lowxigrowthion}] are shown in dotted purple, with a further limit of large $a/\rho_{Li}$ in dashed black. The large $\xi$ limit [\eq{highxigrowthion}] is shown in dotted black. 
    The red curve is the numerical solution of the magnetized Weibel instability, for LHCP (ion cyclotron branch), i.e., using \eq{full-disp} with the magnetized definition of $\xi_\alpha$ [\eq{altxi}] and including the term in \eq{magnetterm}. The orange curve is for RHCP (whistler branch).
    Yellow circles denote linear growth rates extracted from simulations, while red circles correspond to the nonlinear growth rates. 
    }
\label{fig:growth_rate_CS_thickness}
\end{figure}

\begin{figure}
    \centering
     \includegraphics[width=0.9\linewidth]{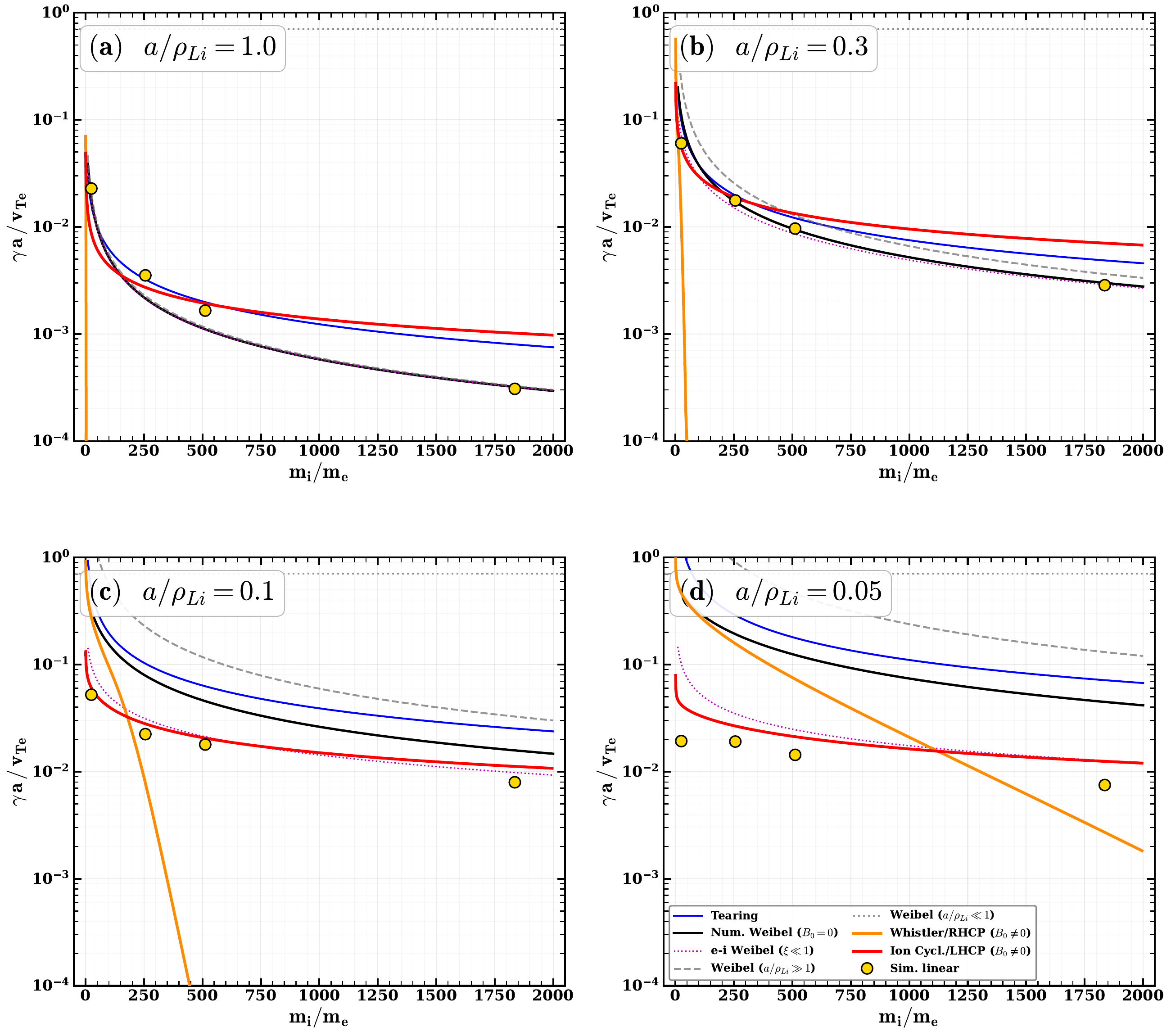}
    \caption{Normalized growth rates $\gamma a / v_{Te}$ as a function of mass ratio $m_i/m_e$ for four current sheet half-thicknesses: (a) $a/\rho_{Li} = 1.0$, (b) $a/\rho_{Li} = 0.3$, (c) $a/\rho_{Li} = 0.1$, and (d) $a/\rho_{Li} = 0.05$. 
    The simulation measurements and theoretical curves are the same as presented in Fig.~\ref{fig:growth_rate_CS_thickness}.}
    \label{fig4}
\end{figure}
Figure~\ref{fig4} presents the same comparison as
Fig.~\ref{fig:growth_rate_CS_thickness}, now plotted as a function of mass
ratio $m_i/m_e$ for four fixed current sheet half-thicknesses:
$a/\rho_{Li} = 1.0$, $0.3$, $0.1$, and $0.05$. For the thickest sheet
[panel~(a)], the simulation growth rates decrease with increasing mass ratio and closely follow both the tearing prediction and the ion cyclotron/LHCP branch.  For thinner current sheets [panels~(b)--(d)], the simulation points
progressively depart from the tearing curve and are best described by the ion
cyclotron branch across the full range of mass ratios. The black dotted line marks the value from \eq{highxigrowthion}; all
simulation points remain well below this threshold.
\begin{figure}
    \centering
        \includegraphics[width=\textwidth]{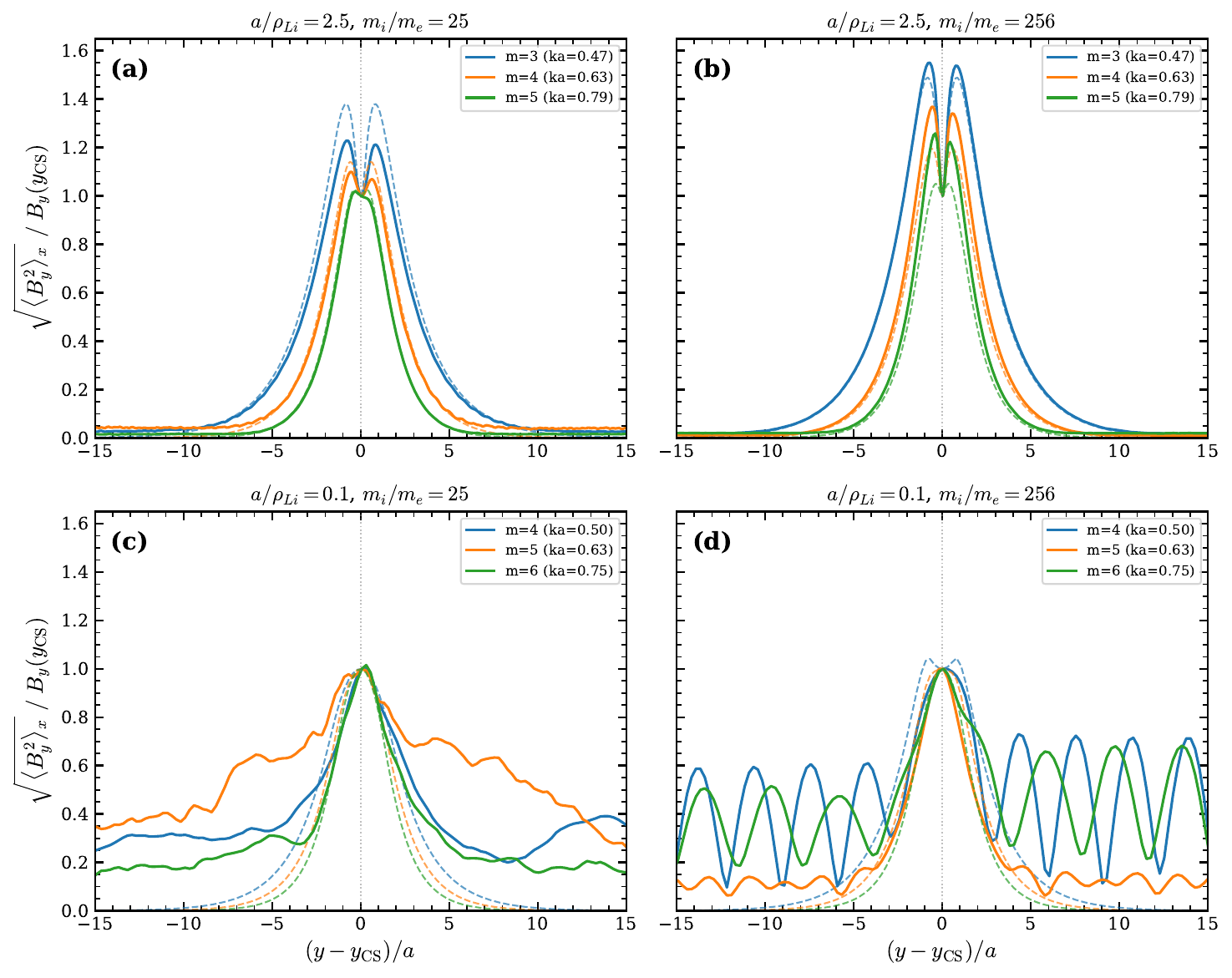}
    \caption{Eigenfunction profiles of individual Fourier modes of $B_y$
    for current sheets with $a/\rho_{Li} = 2.5$ (top row) and
    $a/\rho_{Li} = 0.1$ (bottom row), at mass ratios $m_i/m_e = 25$
    (left column) and $m_i/m_e = 256$ (right column). Each solid curve
    shows the root-mean-square amplitude
    $\sqrt{\langle B_{y,m}^2 \rangle_{x,t}}$ of a single mode~$m$, averaged
    along~$x$ and over all timesteps within the linear growth phase,
    then normalized by its value at the center of the bottom current sheet~$y_{\rm CS} = L_y/4$ (i.e. $y_1$), $B_{norm} \equiv \sqrt{\langle B_{y,m}^2 (y_{CS})\rangle_{x,t}}$.
    The $m = 3,4,5$, and $6$ modes, closest to the predicted fastest-growing wavenumber
    ($ka \approx 0.57$) are shown in each panel.
    The theoretical outer-layer solutions for each mode are plotted in dashed curves.}
    \label{fig3_By_profile}
\end{figure}

Figure~\ref{fig3_By_profile} shows the transverse structure of the magnetic
perturbation, for a select few Fourier modes in the $x$ direction ($m=3,4,5,$ and $6$, close to the fastest growing mode $ka = 2\pi m a/L_x \approx 0.5$.), quantified through 
$\sqrt{\langle B_{y,m}^2\rangle_{x,t}}$, as a function of the normalized distance
from the center of the first current-sheet $\bar{y} \equiv (y-y_{1})/a$, where $y_1 = L_y/4$, for different sheet
thicknesses and mass ratios. The average of $B_{y,m}^2$, where we have isolated a particular mode $m$ in $B_{y}$ using a discrete Fourier transform, labeled as $B_{y,m}$, is performed over $x$ and all time steps within the linear growth phase.

For the thicker current sheet, $a/\rho_{Li}=2.5$,
shown in panels~(a) and~(b), the perturbations are strongly localized around
the sheet center and decay rapidly away from it. Dominant modes exhibit a dip
at the center of the current sheet with peaks to either side, the
characteristic eigenfunction signature of the tearing instability. 
This dip in the eigenfunction corresponds to a positive value of the tearing stability parameter $\Delta^\prime$, given by the change in slope of the eigenfunction in the inner scale \citep{CassakThesis}. The classical tearing instability is stable if this value is negative.
Increasing
the mass ratio from $m_i/m_e=25$ to $256$ further sharpens the inner scale,
producing a narrower and more intense discontinuity around the current sheet center. The $B_y$ profiles match remarkably well with the theoretical curve of the outer solution for a Harris setup, $B_y(\bar{y}) \propto \exp\left(-ka|\bar{y}|\right)\left[1 + (1/ka) \tanh\left(|\bar{y}|\right)\right]$~\citep{CassakThesis}, which is also plotted in~\ref{fig3_By_profile} in dashed lines. To make the fit slightly more accurate, we match the outer solution to a quadratic inner solution (for $|\bar{y}| < \bar{y}_0$), $B_y(\bar{y}) = B_{y0}\left[1+(B_{y,bound}/B_{y0}-1)(\bar{y}/\bar{y}_0)^2\right]$, such that $B_y(\pm \bar{y}_0) = B_{y,bound}$ and $dB_y(\pm \bar{y}_0)/d\bar{y}$ are both matched at the boundaries $\bar{y}=\pm \bar{y}_0$. The matching radius
$\bar{y}_0$ is set to the geometric mean of the electron gyroscale and
the sheet half-thickness,
\begin{equation}
  \bar{y}_0 \;=\; \sqrt{\frac{\rho_{Le} a}{a^2}}
  \;=\; \left(\frac{a}{\rho_{Li}}\right)^{-1/2}\left(\frac{m_i}{m_e}\right)^{-1/4}\,,
  \label{eq:cutoff}
\end{equation}
consistent with the matching point from \cite{Laval1966} and \cite{Coppi1966}.
Because $\bar{y}_0$ depends on 
the mass ratio and thickness, for the
cases in Fig.~\ref{fig3_By_profile} it ranges from
$\bar{y}_0\approx0.16$ ($m_i/m_e=256$, $a/\rho_{Li}=2.5$) to
$\bar{y}_0\approx1.4$ ($m_i/m_e=25$, $a/\rho_{Li}=0.1$).
Note that the assumed matching point $\bar{y}_0$ from \eq{eq:cutoff} is only strictly valid for regimes where $a/\rho_{Le} \gg 1$.

In contrast, for the very thin current sheet $a/\rho_{Li}=0.1$, shown in panels~(c) and~(d), the mode structure changes qualitatively. The dip at the center associated with tearing vanishes. The classical tearing instability should therefore be stable in this regime. The instability we observe, therefore, must be driven by another free-energy source; for example, the effective temperature anisotropy. The disappearance of the dip is consistent with the departure of the growth rate from the tearing prediction observed in
Figs.~\ref{fig:growth_rate_CS_thickness} and~\ref{fig4}.

There appears to be a transition from a classical tearing regime at thick current sheets, where $\Delta^\prime$ drives the instability, and the thin current sheet regime, where Weibel physics likely plays the essential role and effective anisotropy drives the instability.

\section{Conclusion}\label{sec:conclusion}
Although we do not claim that the tearing instability is the same as the Weibel instability, we propose that for thin current sheets, the unstable mode commonly identified as the tearing instability is better described by the Weibel instability than the classic tearing instability. A comparison of the linear theories has revealed that the two instabilities are deeply analogous. Specifically, the Harris configuration can be framed in terms of the Weibel instability by treating the electron drift within the current sheet as an effective temperature anisotropy. Although the Weibel and the tearing modes have similar eigenvectors (and growth rates for $a/\rho_{Li} \sim 1$), for thick current sheets, the growth rates scale differently with current thickness. Consequently, Weibel effects are negligible in thick configurations but become essential in thin current sheets, where classical tearing assumptions break down.

We have tested a range of current sheet thicknesses and mass ratios, using semi-implicit PIC simulations.
For thick current sheets where $a/\rho_{Li} \gg 1$, the classical tearing instability gives a good prediction for the growth rate, as we have shown for the cases with $a/\rho_{Li}=2.5$; the theoretical growth rate of the Weibel instability is small enough that it can be ignored in this regime.
For thin current sheets with $a/\rho_{Li} < 1$, we find that the classical stability parameter $\Delta^\prime$, is negative and thus the classic tearing instability should be stable. Consequently, the "tearing" mode, routinely described in this regime in kinetic studies, is better described by Weibel physics.

The growth rate of moderately thick current sheets ($a/\rho_{Li} \approx 1$) can be estimated by both the Weibel and tearing models. We show, however, that the magnetized version of the Weibel instability, which corresponds to the ion-cyclotron instability, accurately predicts the growth rate of the instability for a wide range of thicknesses ($a/\rho_{Li} \leq 1$). This is surprising, as we made the crude approximation of a constant uniform magnetic field and have ignored the more complicated magnetic field profile of a Harris configuration.

Although we did not explicitly include standard temperature anisotropies, we have demonstrated the intimate relation of the tearing instability for thin current sheets with effective anisotropies in the particle distributions. Consequently, a prediction of how the instability depends on standard temperature anisotropies emerges naturally from this framework and can be further verified in future studies.
Although effects based on ion-to-electron temperature ratios are also evident in the models, a proper study of this parameter using kinetic simulations is left for future work.
Furthermore, we have assumed, in all cases, that the magnetic fields are antiparallel. Studies including a guide field will also be considered in future work.


\section*{Funding} 
This work is supported by the German Science Foundation DFG within the Collaborative Research Center SFB1491.  MEI acknowledges support from the Deutsche Forschungsgemeinschaft (German Research Foundation), grant number 497938371. H.A. acknowledges support from the DFG, grant number 544893192. The authors gratefully acknowledge the Gauss Centre for Supercomputing (GCS) e.V. (www.gauss-center.eu) for funding this project by providing computing time on the GCS Supercomputer SuperMUC-NG at Leibniz Supercomputing Centre (www.lrz.de) through the projects “Heat flux regulation by collisionless processes in heliospheric plasmas—ARIEL” and "Investigation of suprathermal features in the velocity distribution functions of space and astrophysical plasmas-SupraSpace". We gratefully acknowledge the computational resources and scientific support
provided by the HPC cluster Elysium of Ruhr University Bochum, on which some of the particle-in-cell simulations in this study were performed.

\section*{Declaration of interests}
The authors report no conflict of interest.
\section*{Data availability statement}
The main data and input files that support the findings of this study are openly available in Zenodo 

\section*{Author orcid}
K. M. Schoeffler,
https://orcid.org/0000-0002-7644-7381;
H. Aravindakshan,
https://orcid.org/0000-0003-4663-4255;
F. Pucci,
https://orcid.org/0000-0003-4161-8512;
M. E. Innocenti,
https://orcid.org/0000-0002-5782-0013

\appendix
\section{electron Weibel instability}
Here we consider a simple case of the Weibel instability with two species: an electron population and an ion population, which cancels the charge and current of the electrons, but is considered immobile.

We can then simplify the dispersion relation \eq{full-disp} to
\begin{equation}
k^2c^2 -\omega^2 + \omega_{pe}^2 + \omega_{pe}^2 \left(A_e + 1 + 2M_{ez}^2\right) W\left(\xi_e\right)=0
\end{equation}

The Weibel instability is divided into a fast non-resonant instability ($\xi_e \gg 1$), normally occurring when the anisotropy is much greater than unity, and a resonant instability ($\xi_e \ll 1$), where the anisotropy is normally considered small.

\subsection{small $\xi_e$ limit}
In the limit of small $\xi_e$, one can solve for a growth rate.
\begin{equation}
\label{lowxigrowth}
\gamma\left(k\right) = k v_{Te\perp}
\left(\frac{1}{\pi}\right)^{1/2}
\frac{1}{1 + A_e + 2M_{ez}^2} 
\left(A_e + 2M_{ez}^2-k^2d_e^2\right)
\end{equation}
with a peak growth rate at
\begin{equation}
\label{lowxikmaxelec}
kd_e = \sqrt{\frac{A_e + 2M_{ez}^2}{3}} 
\end{equation}
Note that an identical equation is obtained if you include a second species of electrons with equal and opposite drift velocity, rather than the ion species. 
When the drift between the two electron species is small compared to the temperature, the thermal Weibel instability looks essentially the same as the current filamentation instability.
If one defines a temperature $T_{ez,\rm{eff}} \equiv \int dv_z m_e v_z^2f_e(v_z)/ \int dv_z f_e(v_z)$, and then calculates the integral of the sum of the two drifting Maxwellian distributions, one obtains $T_{ez,\rm{eff}} = T_{ez}(1 + 2 M_{ez}^2)$.
Therefore, if $T_{ez} = T_{e\perp}$, the effective anisotropy is thus  
\begin{equation}
\label{AeffApp}
A_{e,\rm{eff}} \equiv \frac{T_{ez,\rm{eff}}}{T_{e\perp}} - 1 = 2M_{ez}^2.  
\end{equation}
The current filamentation instability driven by $M_{ez}$, and the thermal Weibel instability driven by $A_e$ are thus similar and have equivalent growth rates.

Given that you can lump $A_e$ and $M_{ez}$ as one term, we will now consider the instability driven by $M_{ez}$, and assume $A_e=0$ to simplify the equations. Also, considering $\xi_{e} \ll 1$, we assume $M_{ez} \ll 1$ and the fastest growing mode grows at 
\begin{equation}
\label{lowxigrowthfmode}
\gamma = \gamma_0 \equiv  2\omega_{pe}\frac{v_{Te}}{c}\left(\frac{2}{3}\right)^{3/2}
\left(\frac{1}{\pi}\right)^{1/2}
M_{ez}^3
\end{equation}

\subsection{large $\xi_e$ limit}
In the limit $\xi_e\gg 1$, one can solve for a growth rate.

\begin{equation}
\label{hixigrowth}
\gamma\left(k\right) = k v_{Te}\frac{1}{\sqrt{2}}
\sqrt{\frac{2M_{ez}^2}{1+k^2d_e^2}} = 
\frac{k v_{de}}{\sqrt{1+k^2d_e^2}}
\end{equation}
For large wavenumbers, the growth rate is $\gamma= \omega_{pe} v_{de}/c$, independent of $k$. If one includes second-order terms of the $\xi_{e} \gg 1$ approximation, one can find a peak growth rate at
\begin{equation}
\label{hixikmax}
kd_e =\left(\frac{2M_{ez}^2}{3}
\right)^{1/4} 
\end{equation}
Note that the large-wavenumber assumption $k^2d_e^2 \gg 1$ is valid when $\xi_{e} \gg 1$ because this implies $M_{ez} \gg 1$.

The $\xi_e\gg 1$ limit of the Weibel instability also looks the same as the current filamentation instability mathematically; only the velocity distribution function now looks different with a clear two-peak structure in contrast with the thermal Weibel instability, different from the small-$\xi_e$ limit. 

\subsection{electron--ion Weibel}
Next, if we include the effects of the ion motion, but continue to ignore the anisotropy, we can write the dispersion relation as
\begin{equation}
k^2c^2 -\omega^2 + \omega_{pe}^2 + \omega_{pi}^2 + \omega_{pe}^2 \left(1 + 2M_{ez}^2\right) W\left(\xi_e\right)
+ \omega_{pi}^2 \left(1 + 2M_{iz}^2\right) W\left(\xi_i\right)=0
\end{equation}
or
\begin{equation}
\label{iedispersion}
k^2c^2 -\omega^2 - \omega_{pe}^2 2M_{ez}^2 - \omega_{pi}^2 2M_{iz}^2- \omega_{pe}^2\left(1 + 2M_{ez}^2\right) \xi_e Z\left(\xi_e\right)
- \omega_{pi}^2 \left(1 + 2M_{iz}^2\right) \xi_i Z\left(\xi_i\right)=0
\end{equation}
\subsection{small $\xi_e,~\xi_i$ limit}
Assuming $v_{de}/T_e = v_{di}/T_i$, as is required in a Harris equilibrium, we can solve \eq{iedispersion} for a growth rate in the limit of small $\xi_e$ and $\xi_i$
\begin{equation}
\label{lowxigrowthion}
\gamma\left(k\right) = k v_{Te,\rm{eff}}
\left(\frac{1}{\pi}\right)^{1/2}
\left(2M_{ez}^2\eta_e-k^2d_{e}^2\right)
\end{equation}
with a peak growth rate at
\begin{equation}
\label{lowxikmaxionApp}
kd_e = \sqrt{\frac{2M_{ez}^2\eta_e}{3}} 
\end{equation}
where $v_{Te,\rm{eff}}$ is an effective thermal velocity
\begin{equation}
v_{Te,\rm{eff}} \equiv \frac{v_{Te}}{\left(1+\left(\mu T_e/T_i\right)^{1/2}\right)\left(1+2M_{ez}^2\left(\mu T_e/T_i\right)^{-1/2}\right)}.
\end{equation}
Note the similar form to \eqs{lowxigrowth}{lowxikmaxelec}.
If we normalize to $a$ assuming a Harris configuration where $a=d_e/\sqrt{M_{ez}^2\eta_e}$ from \eq{classrho}
\begin{equation}
\label{lowxikmaxiona}
k a = \sqrt{\frac{2}{3}} .
\end{equation}

If we assume a low anisotropy, or rather $M_{ez} \ll \left(\mu T_e/T_i\right)^{1/4}$ (we will show that this is a requirement for the small $\xi_e$ and $\xi_i$ approximation), and we normalize to $v_{Te}/a$ again assuming a Harris configuration where $a=d_e/\sqrt{M_{ez}^2\eta_e}$,
we can simplify the growth rate to
\begin{equation}
\label{lowxigrowthionsimp}
\frac{\gamma\left(k\right) a}{v_{Te}} = \frac{kd_e}{\sqrt{M_{ez}^2\eta_e}} 
\left(\frac{1}{\pi}\right)^{1/2}
\frac{1}{1+\left(\mu T_e/T_i\right)^{1/2}} 
\left(2M_{ez}^2\eta_e-k^2d_e^2\right)
\end{equation}
At the fastest growing $k$,
\begin{equation}
\label{lowxigrowthionfmodeApp}
\frac{\gamma a}{v_{Te}} = 2 \left(\frac{2}{3}\right)^{3/2}
\left(\frac{1}{\pi}\right)^{1/2}
\frac{1}{1+\left(\mu T_e/T_i\right)^{1/2}} 
\eta_e
M_{ez}^2
\end{equation}
One should note the similarity in form with the tearing growth rate in \eq{tearingMez}.
One can also verify the assumption of small $\xi_e$ and $\xi_i$,
\begin{equation}
\label{xelowxigrowthionfmode}
|\xi_e| = \frac{\gamma}{k v_{Te}} = 2 \left(\frac{2}{3}\right)
\left(\frac{1}{\pi}\right)^{1/2}
\frac{1}{1+\left(\mu T_e/T_i\right)^{1/2}} 
\eta_e
M_{ez}^2
\end{equation}
\begin{equation}
\label{xilowxigrowthionfmode}
|\xi_i| = \frac{\gamma}{k v_{Ti}} = 2 \left(\frac{2}{3}\right)
\left(\frac{1}{\pi}\right)^{1/2}
\frac{\left(T_e/T_i\right)^{1/2}\mu^{-1/2}}{1+\left(\mu T_e/T_i\right)^{1/2}} 
\eta_e
M_{ez}^2
\end{equation}
which holds as long as both $ (\eta_e M_{ez}^2)^{-1/2} \sim a/d_e $ and $((T_e/\mu T_i)^{1/2}\eta_e M_{ez}^2)^{-1/2} \sim a/d_{e} \cdot (T_i\mu/T_e)^{1/4}$ are much more than unity.
In this study, we will focus on the regime where $T_e = T_i$, so our assumptions are valid for thick current sheets where $a/\sqrt{d_ed_i}>1$.

Now let us consider a thin current sheet, or high anisotropy $M_{ez} \gg \left(\mu T_e/T_i\right)^{1/4}$, but still in the small $\xi_e$ and $\xi_i$ limit.
At the fastest growing $k$, which is unchanged,
\begin{equation}
\label{lowxigrowthionfmodesmalla}
\frac{\gamma a}{v_{Te}} = \left(\frac{2}{3}\right)^{3/2}
\left(\frac{1}{\pi}\right)^{1/2}
\frac{\left(\mu T_e/T_i\right)^{1/2}}{1+\left(\mu T_e/T_i\right)^{1/2}} 
\eta_e.
\end{equation}
In this case,
\begin{equation}
|\xi_i| = \frac{\gamma}{k v_{Ti}} = \left(\frac{2}{3}\right)
\left(\frac{1}{\pi}\right)^{1/2}
\frac{T_e/T_i}{1+\left(\mu T_e/T_i\right)^{1/2}} 
\eta_e,
\end{equation}
which is close to $1$ when $T_e=T_i$, so our assumption $|\xi_i| \ll 1$ breaks down. However, when a magnetic field is included, for the LHCP mode, ion cyclotron resonances limit $|\xi_i|$ to approximately $1$, and this approximation still gives a good order-of-magnitude estimate.

\subsection{large $\xi_e,~\xi_i$ limit}
If we instead assume $|\xi_e|, |\xi_i| \gg 1$, as might be expected for very thin current sheets, we obtain
\begin{equation}
    \gamma(k) = \frac{k v_{Te}}{\sqrt{2}} \left(\frac{ 2M_{ez}^2 \left(1 + \mu T_i^2/ T_e^2\right) + 1 + \mu^2 T_i/T_e}{1 + \mu+ k^2 d_e^2}\right)^{1/2}
\end{equation}
and, like with \eq{hixikmax}, when including higher order terms in the small $1/\xi_e$,$1/\xi_i$ expansion, we find a peak growth rate at
\begin{equation}
    kd_e = \left(\frac{1}{3}\frac{ \left(2M_{ez}^2 \left(1 + \mu T_i^2/ T_e^2\right) + 1 + \mu^2 T_i/T_e\right)^2\left(1+\mu\right)}{2M_{ez}^2 \left(1 + \mu^2 T_i^3/ T_e^3\right) + 1 + \mu^3 T_i^2/T_e^2}\right)^{1/4}
\end{equation}
The asymptotic growth rate for large $k$
and when $M_{ez} \gg 1$ (we will show that this is a requirement for the large $\xi_e$ and $\xi_i$ approximation),
is given by
\begin{equation}
    \label{highxigrowthion}
    \frac{\gamma a}{v_{Te}}  = \left(1 + \mu T_i^2/ T_e^2\right)^{1/2}\eta_{e}^{-1/2}
\end{equation}
and
\begin{equation}
    \label{highxikmaxion}
    k a = (M_{ez}\eta_e)^{-1/2}\left(\frac{2}{3}\frac{ \left(1 + \mu T_i^2/ T_e^2\right)^2\left(1+\mu\right)}{\left(1 + \mu^2 T_i^3/ T_e^3\right)}\right)^{1/4}
    \sim \left(\frac{a}{d_e}\right)^{1/2}\left(\frac{1}{\eta_e}\right)^{1/4}
\end{equation}
One can also verify the assumption of large $\xi_e$ and $\xi_i$,
\begin{equation}
    |\xi_e|  = \frac{\gamma}{k v_{Te}} 
    =\left(\frac{3}{2}\left(1 +  \mu^2T_i^3/ T_e^3\right)/\left(1+\mu\right)\right)^{1/4}
    M_{ez}^{1/2}
\end{equation}
\begin{equation}
    |\xi_i| = \frac{\gamma}{k v_{Ti}}   
    =
    \left(\frac{3}{2}\left(1 +  \mu^2T_i^3/ T_e^3\right)/\left(1+\mu\right)\right)^{1/4}
     \left(\frac{T_e}{T_i}\frac{M_{ez}}{\mu}\right)^{1/2}
\end{equation}
which holds as long as both $M_{ez}^{-1} \sim \eta_e^{1/2}\cdot a/d_e$ and $(M_{ez}T_e /\mu T_i)^{-1} \sim (T_i\mu/T_e)\eta_e^{1/2} \cdot a/d_{e} $ are much less than unity, i.e., for very thin current sheets. Although this growth rate does not match any of our simulations, this may play a role for electron-dominated regimes, where $T_e \gg T_i$ (or non-Harris setups). 

\bibliographystyle{jpp}

\bibliography{Weibeltearing}

\end{document}